\RequirePackage[2020-02-02]{latexrelease}

\documentclass{aes2e}

\usepackage{epstopdf}
\usepackage[caption=false]{subfig}

\usepackage{epstopdf}
\usepackage[bookmarksnumbered,colorlinks,bookmarks,citecolor=blue,linkcolor=blue,urlcolor=blue,breaklinks,linktocpage]{hyperref}
\usepackage[all]{hypcap}
\usepackage{listings}
\usepackage{amsmath}
\usepackage{mathtools}

\usepackage{amsmath} 
\usepackage{graphicx}
\usepackage{url}
\makeatletter
\g@addto@macro{\UrlBreaks}{\UrlOrds}
\makeatother

\usepackage{url}

\begin{document}

\markboth{Lazzarini and Timoney}{}

\title{Improving the Chamberlin Digital State Variable Filter}

\authorgroup{
\author{Victor Lazzarini}
\role{ }
and  \author{Joseph Timoney},
\role{AES Member}
\email{(victor.lazzarini@mu.ie)\quad\quad\quad\quad\quad\quad\quad\quad\quad\quad\quad\quad (Joseph.Timoney@mu.ie)}
\affil{Maynooth University, Maynooth, Ireland}
}

\abstract{%
The state variable filter configuration is a classic analog design which has been employed in many
electronic music applications. A digital implementation of this filter was put forward by Chamberlin, which
has been deployed in both software and hardware forms. While this has proven to be a straightforward
and successful digital filter design, it suffers from some issues, which have already been identified in the
literature. From a modified Chamberlin block diagram, we derive the transfer functions describing its
three basic responses, highpass, bandpass, and lowpass. An analysis of these leads to the development
of an improvement, which attempts to better shape the filter spectrum. From these new transfer functions,
a set of filter equations is developed. Finally, the approach is compared to an alternative time-domain 
based re-organisation of update equations, which is shown to deliver a similar result.
}


\maketitle
\section{Introduction}

Analog filters have been a source of inspiration for digital filter design since the early days of the discipline of digital signal 
processing. This solves one of the great conundrums of the area, particularly for musical applications, which is how to provide
useful filters that are punchy in character (that is, have a powerful sound transformation effect, particularly with regards to
their amplitude responses), have low computational demands, and are easy to design or control. The first two aspects are
easily met by infinite impulse response (IIR), or feedback, filters, while the third is, strictly, only possible with finite impulse
response filters. However, by recourse to the various configurations that were developed for analog filter circuits, it is possible
to bridge that gap and have IIR digital filters that can be deployed in many scenarios.

Therefore it is common practice to look for inspiration in analog signal processing \cite{Rossum}, particularly in the case of celebrated 
designs such as the various ladder filter implementations and their variations \cite{Stilson, Huovilainen, Fontana3, Dangelo1, Dangelo2}. In this paper, we look at another very interesting
analog filter case, which is the state variable filter. This has been employed in many significant musical instrument applications.
 very early on received a digital treatment in the pioneering work
of Chamberlin \cite{Chamberlin}, which has been a source of inspiration for many computer music practitioners, and deployed in very successful
instruments such as the Kurzweil synthesizers \cite{Dattorro}. However, this design has some
practical issues that may turn out to be problematic in some applications. In this paper, we explore the question of improving the design from a spectral perspective. The filter developed here can then be used as a drop-in replacement for the Chamberlin design, or employed to implement
a variety of responses such as the Octave Cat and Moog ladder lowpass filters as described by \cite{WernerMcClellan}.

The text is organised as follows. We first introduce the original analog state variable configuration and
examine its characteristics. Next we turn to the Chamberlin implementation, deriving an equivalent biquadratic transfer function to
describe its spectrum. This leads into an analysis of the issues stemming from it. We carry this spectral approach further and propose an improved frequency response, which then informs some modifications to the digital filter update equations. Finally,
we discuss the method and contrasting approaches leading to similar results. Csound language code is used as ready-to-deploy
examples of the filter designs discussed in the paper.

\section{The State Variable Filter}

The \emph{state variable filter} \cite{Kerwin1967StateVariableSF, Colin1971ElectricalDA, Hutchins2} is a classic analog filter configuration, which has been employed in numerous 
musical applications. A typical example is found in the second-order lowpass filter of the Oberheim OB-Xa synthesiser, which uses
a CEM3320 integrated circuit to realize it \cite{ElectricDruid}. The state variable filter can be described as an integrator-based design, which sets it apart from the leaky-integrator forms found in typical first-order lowpass sections of ladder filters. Another special aspect of the filter, which was somehow ignored in the Oberheim implementation, is that it can provide a whole variety of frequency responses, lowpass, highpass, bandpass, band-reject, and allpass, simultaneously. In addition, Hutchins \cite{Hutchins143} demonstrated that two state variable filters in series can be used to implement a frequency response similar to the ladder filter, which was again shown in \cite{WernerMcClellan}.

\begin{figure}[htp]
\begin{center}
\includegraphics[width=.4\textwidth]{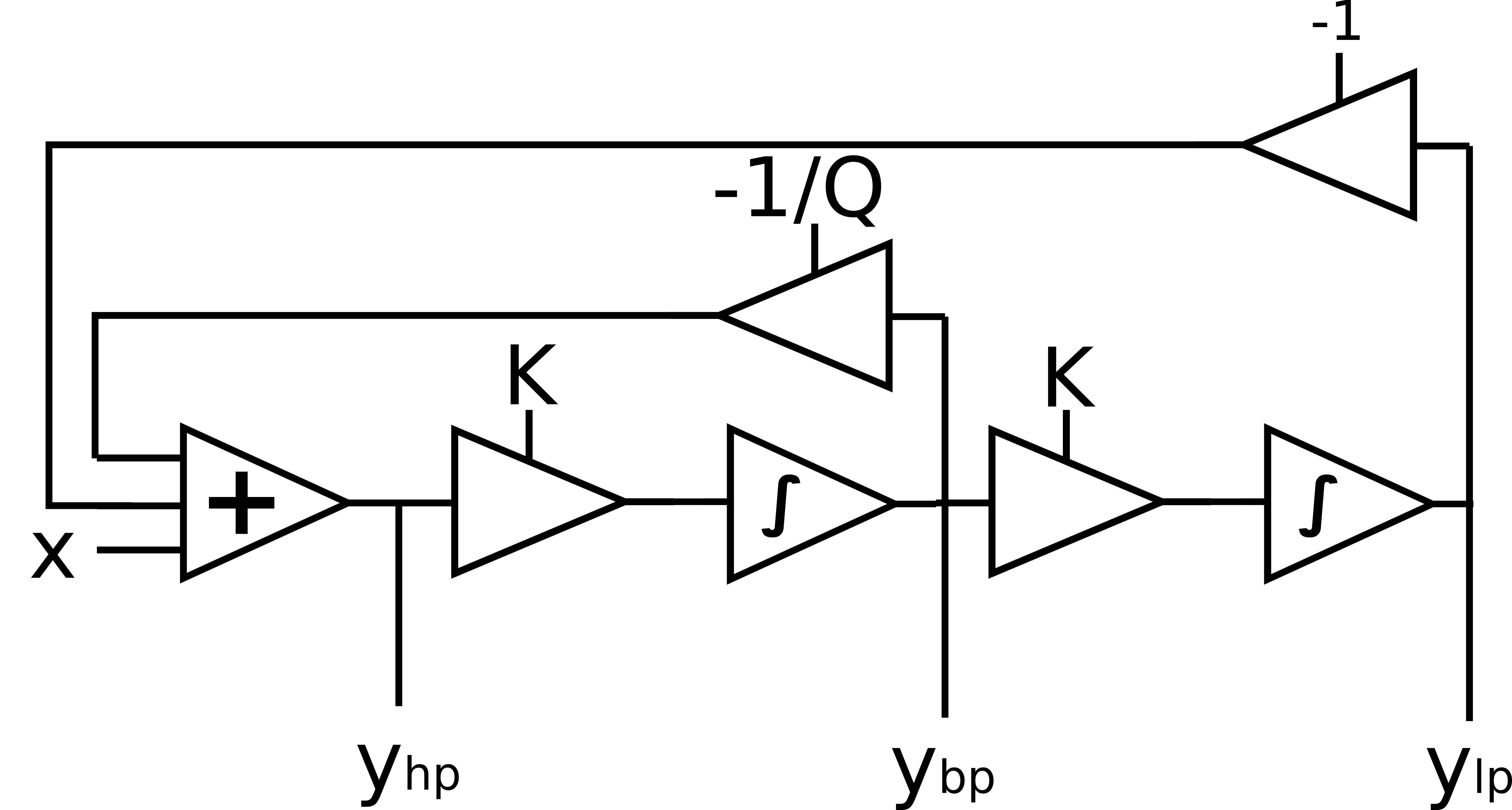}
\caption{State variable filter block diagram.}
\label{fig:statevar}       
\end{center}
\end{figure}

The block diagram of the state variable filter is shown in Fig.~\ref{fig:statevar}, where we observe that it is composed of two
integrators in series, both of which are fed back to be summed with the filter input, scaled by $-1/Q$ and $-1$, respectively. From this, two parameters control the frequency
response, $Q$ and $K$. The latter is proportional to the filter frequency, whereas the former determines the amount of resonance
or peaking around that frequency. To get a highpass output, we tap the filter after the input summing stage. The bandpass signal
comes from the first-order integrator output, and the second-order integrator output gives the lowpass response. A
system of filter equations can then be defined as

\begin{equation}
\begin{split}
&y_{hp}(t) =  x(t) - (1/Q) y_{bp}(t)  - y_{lp}(t) \\
&y_{bp}(t) = K\int_0^t  y_{hp}(\theta) d\theta \\
&y_{lp}(t) = K\int_0^t y_{bp}(\theta) d\theta .
\end{split}
\end{equation}
\smallskip

Due to the two integrators in series, the highpass and lowpass outputs have a phase difference of $\pi$ radians at the cutoff frequency, therefore by mixing them together, we can obtain a band-reject response. Summing the three outputs yields an allpass response. As far as analog filters are concerned, the state variable filter has a very straightforward block diagram, with just
three black box components: the integrator, a variable gain element, and a summing unit. While there are various
ways to implement it, the circuits often only require three or four operation amplifiers, plus a few resistors and capacitors.

The transfer functions for each output can be derived as follows. First we define the general lines
of the highpass transfer function as

\begin{equation}\label{eq:svar_an}
H_{hp}(s) = 1 - (1/Q) H_{bp}(s) - H_{lp}(s).
\end{equation}
\smallskip

\noindent Then, using the integrator transfer function $H(s) = 1/s$, we have

\begin{equation}
\begin{split}
&H_{bp}(s) = \frac {K H_{hp}(s)}s \\
&H_{lp}(s)  = \frac {K H_{bp}(s)}s = \frac {K^2 H_{hp}(s)} {s^2} .
\end{split}
\end{equation}
\smallskip

Substituting this equation in Eq.~\ref{eq:svar_an} gives us the highpass transfer function
as

\begin{equation}\label{eq:svar_anhp2}
H_{hp}(s) =  \frac {s^2} {s^2 + (K/Q)s + K^2},
\end{equation}
\smallskip

\noindent followed by the bandpass

\begin{equation}\label{eq:svar_anbp2}
H_{bp}(s) =  \frac {Ks} {s^2 + (K/Q)s + K^2},
\end{equation}
\smallskip

\noindent and lowpass response

\begin{equation}\label{eq:svar_anlp2}
H_{lp}(s) =  \frac {K^2} {s^2 + (K/Q)s + K^2}.
\end{equation}
\smallskip

These three equations allow us to note that the state variable filter is a second-order, or two-pole, 
filter with a biquadratic transfer function. The different frequency responses share the same 
denominator, defining the filter poles, and diverge in the position of the zeros. These are 
two at s = 0, for the highpass filter; one at s = 0 and another at $s = \infty$ for bandpass;
and finally two zeros at $s = \infty$ in the lowpass case.

\section{Chamberlin's Digital State Variable Filter} 

\begin{figure}[htp]
\begin{center}
\includegraphics[width=.2\textwidth]{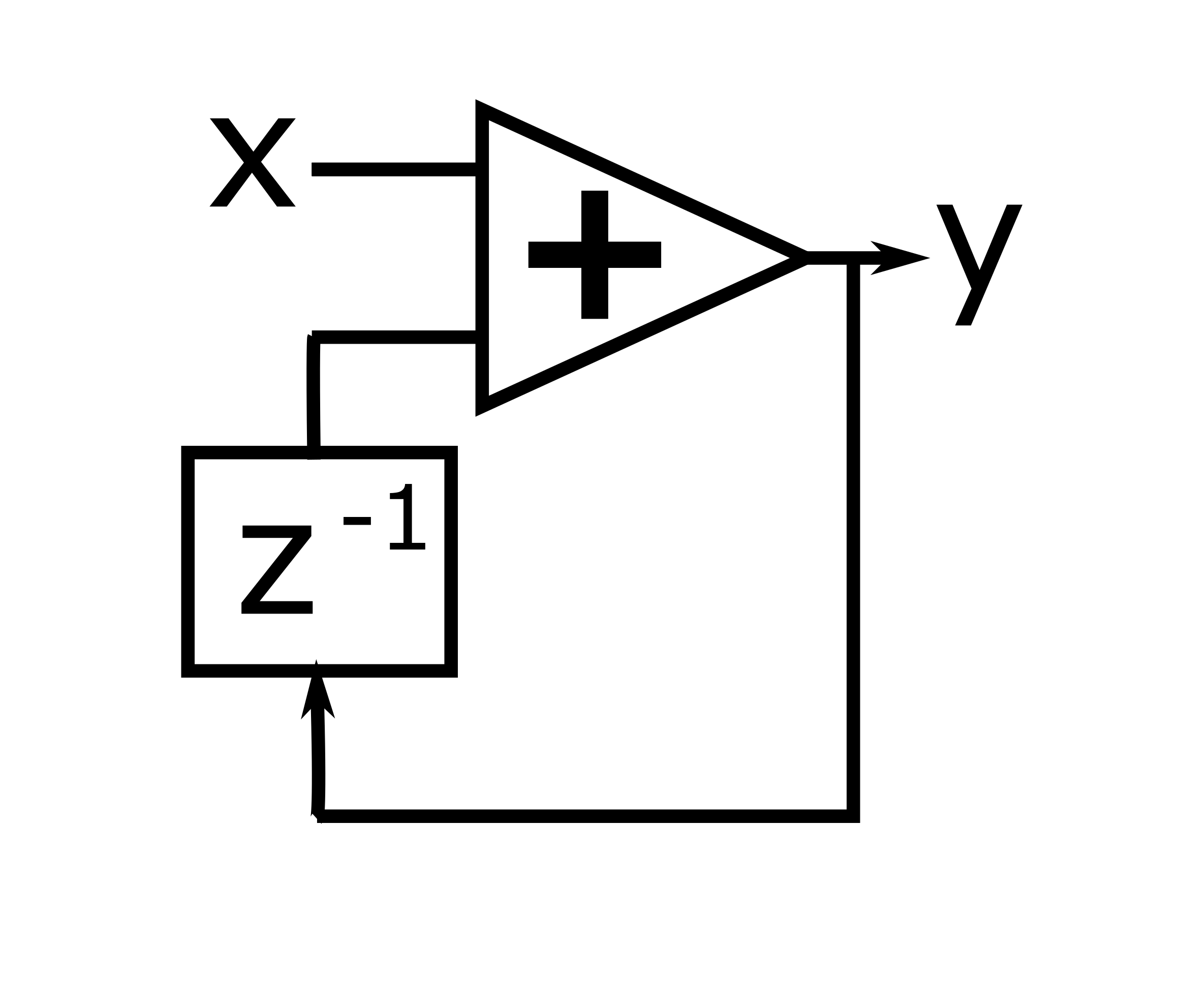}
\caption{The digital integrator.}
\label{fig:integrator}       
\end{center}
\end{figure}

In his book, Chamberlin \cite{Chamberlin} proposes that the state variable filter may be a very economical model for the implementation
of a digital filter for musical applications. The idea is that with only a few operations we would have a processor that is
capable of several different frequency responses, controlled by only two parameters. Moreover, as we noted above,
the filter is of a fairly straightforward design, which also simplifies implementation. The main component is the
integrator, for which a discrete-time version is given in Fig.~\ref{fig:integrator}. This is simply a backward Euler method discretization of continuous time integration, which yields an allpole integrator. Its transfer function is

\begin{equation}\label{eq:integtf}
H(z) = \frac 1 {1 - z^{-1}} .
\end{equation}
\smallskip

\begin{figure}[htp]
\begin{center}
\includegraphics[width=.4\textwidth]{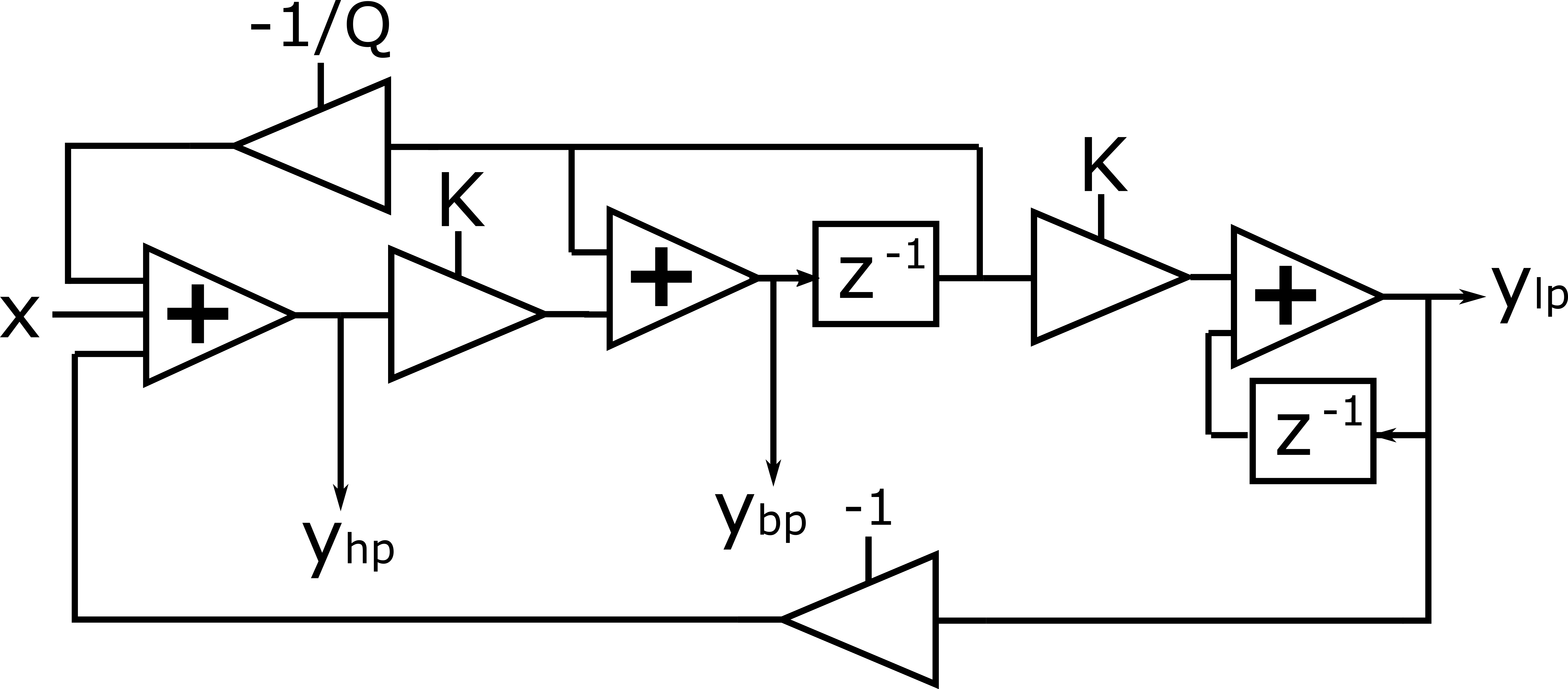}
\caption{Chamberlin digital state variable filter block diagram.}
\label{fig:chamberlin}       
\end{center}
\end{figure}

Chamberlin's design employs the two digital integrators in such a way that yields a slightly modified
block diagram. The change is made at the first integrator stage. Instead of taking the integrator 
output as the input to the next stage, the integrator state (its delay) is tapped to provide this signal. This
inserts a 1-sample delay in the middle of the block diagram (Fig.~\ref{fig:chamberlin}). The idea behind this is to keep 
the phase difference between the highpass and lowpass outputs as in the original analog
model, so that we can obtain a band-reject output. This also requires us to re-order the 
computation so that the lowpass output is computed first, using the following equations 

\begin{equation}
\begin{split}
&y_{lp}(n) = Ky_{bp}(n-1) + y_{lp}(n-1) \\
&y_{hp}(n) =  x(n) - (1/Q)y_{bp}(n-1) - y_{lp}(n) \\
&y_{bp}(n) = Ky_{hp}(n) + y_{bp}(n-1) ,
\end{split}
\end{equation}
\smallskip

\noindent which are implemented in Listing~\ref{code:svar1}.\\

\begin{lstlisting}[caption={Chamberlin digital state variable filter.},label={code:svar1}]
opcode Svar,aaaa,akk
 setksmps 1
 abp,alp init 0,0
 as,kK,kQ xin
 alp = abp*kK + alp 
 ahp = as - alp - (1/kQ)*abp
 abp = ahp*kK + abp
  xout ahp,alp,abp,ahp+alp
endop
\end{lstlisting}

\begin{figure}[htp]
\begin{center}
\includegraphics[width=.4\textwidth]{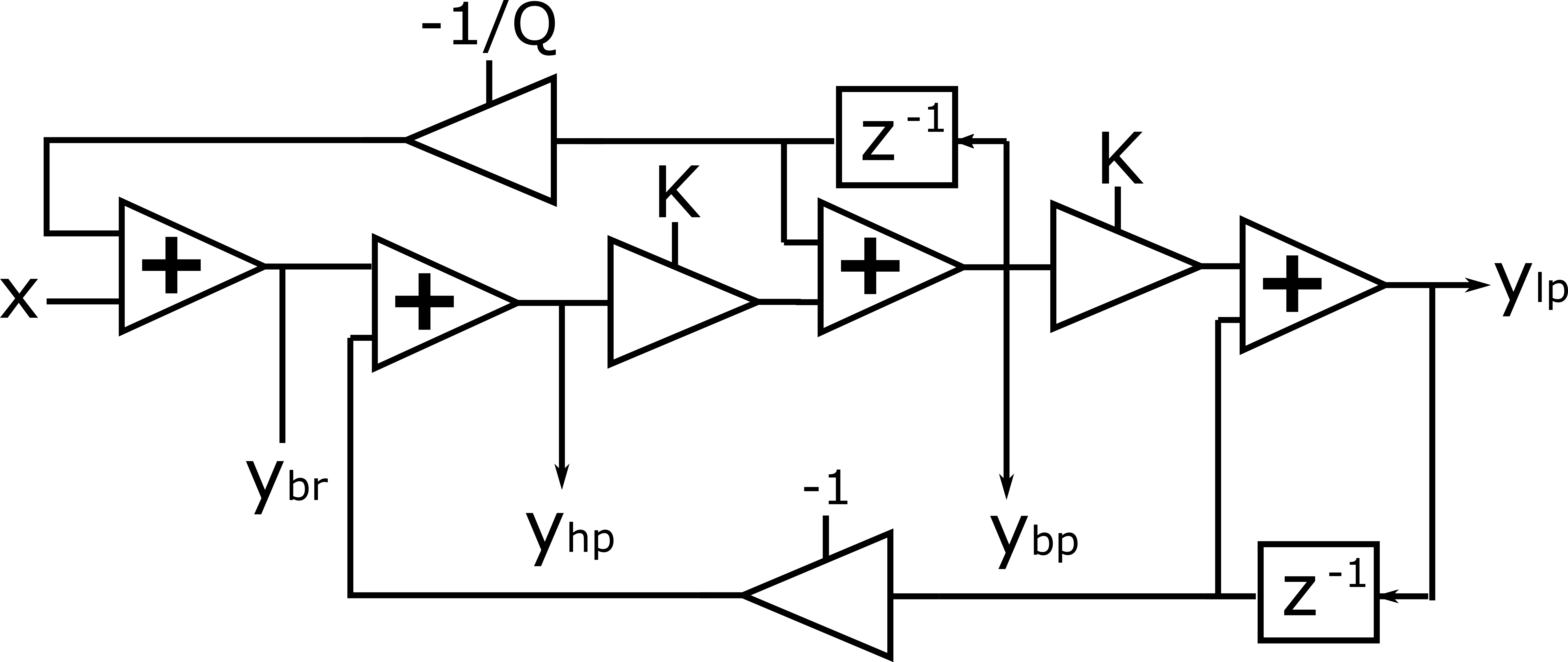}
\caption{Re-arranged Chamberlin digital state variable filter block diagram.}
\label{fig:statevarV}       
\end{center}
\end{figure}

To begin an analysis of this implementation, we might want first to re-arrange this design
into a filter which resembles the original block diagram more closely (Fig.~\ref{fig:statevarV}). 
While we could derive a transfer function for the current arrangement, as indeed \cite{Dattorro} did
for the lowpass response, a configuration that is closer to the analog filter will allow us
to develop the proposed improvements in a more straightforward way. For this, all is needed
is to move the 1-sample delay to both feedback paths and restoring the signal path
through the filter. This is equivalent to tapping the integrator states to get the inputs 
to the highpass filter, as shown by

\begin{equation}
\begin{split}
&y_{hp}(n) =  x(n) - (1/Q)y_{bp}(n-1) - y_{lp}(n-1) \\
&y_{bp}(n) = Ky_{hp} + y_{bp}(n-1) \\
&y_{lp}(n) = Ky_{bp}(n) + y_{lp}(n-1).
\end{split}
\end{equation}
\smallskip

This form produces a highpass and bandpass outputs that are sample-by-sample 
equivalent to those of the Chamberlin filter, and a lowpass output that is also 
identical when delayed by one sample. The main difference between the two 
arrangements is that now  there is a 1-sample delay between the lowpass and bandpass 
signals and the highpass output. Therefore, the phase 
relationship between the lowpass and highpass needed to obtain a band-reject 
response is lost. However, an equivalent relationship can be used for this purpose,
which happens between the input and the inverted bandpass feedback path.
To get the band-reject output we split the sum that produces the highpass signal
in two separate stages, the first of which is equivalent to the desired output. An
examination of the original block chart shows that this is exactly equivalent to
the sum of the highpass and lowpass signals. This rearrangement is shown in 
Listing~\ref{code:svar2}.\\

\begin{lstlisting}[caption={Re-arranged Chamberlin digital state variable filter.},label={code:svar2}]
opcode Svar,aaaa,akk
 setksmps 1
 abp,alp init 0,0
 as,kK,kQ xin
 abr = as - (1/kQ)*abp 
 ahp = abr - alp
 abp = ahp*kK + abp
 alp = abp*kK + alp
     xout ahp,alp,abp,abr
endop
\end{lstlisting}

\subsection{Transfer Functions}

This puts the filter in a form for which we can derive transfer functions as we did in the
analog case. Starting again with an outline of the highpass transfer function, we have

\begin{equation}\label{eq:svar_hp}
H_{hp}(z) = 1 - \frac {z^{-1}} Q H_{bp}(z) - z^{-1}H_{lp}(z),  
\end{equation}
\smallskip

\noindent where $H_{bp}(z)$ and $H_{lp}(z)$ are the bandpass and lowpass responses, respectively. Note that the $z^{-1}$ factor needs to be incorporated to take account of the 1-sample delay in the feedback path, which does not exist in the analog filter. From this, as before, we can derive the bandpass and lowpass responses in turn, as they correspond first- and second-order integration of the highpass output signal, each scaled by a $K$ factor,

\begin{equation}
\begin{split}
&H_{bp}(z) = \frac {K H_{hp}(z)} {1 - z^{-1}} \\
&H_{lp}(z)  = \frac {K H_{bp}(z)} {1 - z^{-1}} = \frac {K^2 H_{hp}(z)} {(1 - z^{-1})^2}.
\end{split}
\end{equation}
\smallskip

Now we can replace these back into Eq.~\ref{eq:svar_hp}, to obtain the correct frequency response for the highpass output

\begin{equation}\label{eq:svar_hp2}
H_{hp}(z) =  \frac {(1 - z^{-1})^2} {1 - (2  - K/Q - K^2) z^{-1} + (1 - K/Q) z^{-2}},
\end{equation}\smallskip

\noindent and from $H_{hp}(z)$  we get 

\begin{equation}\label{eq:svar_bp2}
H_{bp}(z) = \frac {K(1 - z^{-1})} {1 - (2  - K/Q - K^2) z^{-1} + (1 - K/Q) z^{-2}}, 
\end{equation}
\smallskip

\noindent and

\begin{equation}\label{eq:svar_lp2}
H_{lp}(z)  =  \frac {K^2} {1 - (2  - K/Q - K^2) z^{-1} + (1 - K/Q) z^{-2}}.
\end{equation}
\smallskip

These equations do indeed give us a highpass, bandpass, and lowpass filters, with resonance controlled
by the $Q$ parameter, which turns out to be, on first looks, very similar here to its usual interpretation as the ratio 
between frequency and bandwidth.  Now we need to determine how to compute $K$.  We have noted that in
the analog filter, $Q$ is its quality factor, related to resonance, and $K$ is proportional to 
the cutoff/centre frequency. In this case, we can set $K = 2\pi f$, but in this digital model, an
equivalent expression such as  $K = 2\pi f/f_s$ will not be accurate. This is mainly because the digital 
integrators introduce a certain amount of error, particularly as the frequency increases.
A correction factor can be applied, yielding the expression $K = 2\sin(\pi f/f_s)$, 
which gives a more accurate tuning of the filter frequency \cite{Chamberlin}.

\subsection{Issues}

However, some difficulties still remain. If we examine the transfer functions, we will note
that at high frequencies (particularly $> f_s/4$) pole frequency will drift higher than expected 
depending on the value of $Q$. In order to keep the filter more or less in tune at these frequencies, 
we need to increase $Q$. The coupling of the two parameters is evident from
the fact that the pole radius is dependent on both $Q$ and $K$. As $K$ gets
larger, for low values of $Q$, it will make the pole move slightly away from the unit circle. 
Moreover, the pole frequency will also shift upwards relative to where we would want it to be, 
and the filter also may need some means of output scaling to prevent it from exploding. 
The limits of stability and tuning for the Chamberlin, which depend both on $K$ and $Q$, 
have been determined by Dattorro \cite{Dattorro}, as

\begin{equation}
0 < K < \textrm{min}\left(Q, 2 - \frac 1 Q, 2Q -\frac 1 Q, \frac {-1/Q + \sqrt{8 + (1/Q)^2}} 2 \right),
\end{equation}

\noindent from where we can surmise that the filter will not behave very well 
at high frequencies.

\begin{figure}[htp]
\begin{center}
\includegraphics[width=\columnwidth]{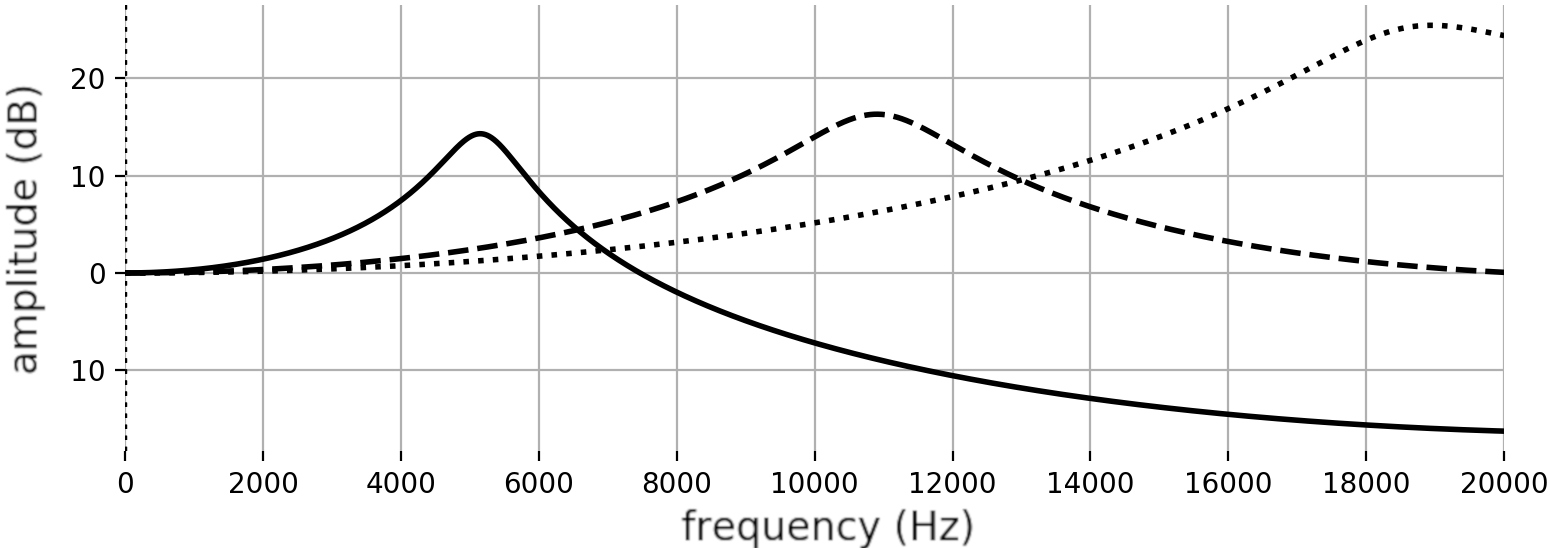}
\caption{Amplitude responses for the state variable lowpass filter, using $Q=5$ and $f=5$ KHz (solid), 10 KHz (dashes), and 15KHz (dots), with $K = 2\sin(\pi f/f_s)$ and $f_s =44.1$ KHz.}
\label{fig:svarerr}      
\end{center}
\end{figure}

Figure~\ref{fig:svarerr} shows these discrepancies in the expected filter frequency and the 
resulting amplitude responses. Notice that these problems are perhaps not themselves an inherent problem 
with the filter, but just a difficulty to find correct values for $Q$ and $K$ to yield the correct filter 
frequency and stability. Running the filter at higher sampling rates will improve the filter 
tuning as the onset of errors is pushed higher in the spectrum.

Perhaps the main reason why the direct translation of the 
state-variable flowchart fails at high frequencies, at least using normal sampling rates, 
is that we had to sneak in a one-sample delay somewhere in the block diagram, because
it is not possible to have an instantaneous-time feedback as in the analog circuit.
If we follow the filter structure, we will notice that the bandpass and lowpass 
outputs are supposed to be the exact same as the inputs to the highpass signal. 
That of course cannot be computed motivating the use of an extra 1-sample delay. 
The problem is that this addition modifies the filter topology somewhat, and the digital version
does not fully match the original analog block diagram in a reasonable way.

We could of course adopt a different route by reverting to the biquadratic analog transfer functions 
in the s-domain, and then apply the bilinear transformation \cite{STEIGLITZ1965} directly to them. 
These will give us coefficients for typical highpass, lowpass, and bandpass second-order 
digital filter sections. This is a particularly useful approach to implement other analog
designs such as the Sallen-Key \cite{Hutchins3} and Steiner-Parker \cite{Steiner} filters, which we have done in \cite{CMJ}.
However, this somehow defeats the purpose of trying to model a 
state variable filter and reap the benefits of this configuration. In particular, this is useful if we want to
expand it into implementations that include nonlinear elements, as we have also shown in \cite{CMJ}.

\section{An Improved Digital State Variable Filter}

While we cannot avoid the fact that a 1-sample delay needs to figure somewhere in
the block diagram, we can improve things by placing it in an optimal position. We can
determine this by looking at the transfer functions and try to establish why they 
are not ideal. Since we identified that the problem appears to be very apparent in
the lowpass case, we have a good place to start. Examining Eq.~\ref{eq:svar_lp2},
we notice that if we moved the two (theoretical) zeros, which exist at $s=\pm\infty$ in the analog 
case, to the $z=-1$, we would improve the lowpass response somewhat. This is not
a complete solution, but gives us a route towards it.

One of the problems of unilaterally fixing the lowpass transfer function is that, if
we are to preserve the state variable structure, this will have to be compensated 
by changes in the other transfer functions. We need to find a way to change the
complete filter so that we end up with two zeros at $z=-1$ in the lowpass filter
frequency response. It is becoming clear that the problem is the complete
absence of zeros at the Nyquist frequency, in both the lowpass and bandpass responses. 
If we look closely at Eq.~\ref{eq:svar_hp}, we will notice that we in fact have two 
actual poles at $z=0$. These are the result of the 1-sample delays we had to 
inflict to the block diagram. We can now try to swap these for zeros in a position where their 
effect can be used to solve the issue we identified in the lowpass response. 

This requires us to replace the pure delays $z^{-1}$ by one-zero lowpass filters $1+z^{-1}$, 
in the Chamberlin highpass transfer function (Eq.~\ref{eq:svar_hp}),

\begin{equation}\label{eq:svar_hp_fix}
H_{hp}(z) = 1 - \frac {(1 + z^{-1})} Q H_{bp}(z) - (1 + z^{-1}) H_{lp}(z), 
\end{equation}
\smallskip

\noindent which places a first-order lowpass finite impulse response filter in each one
of the feedback paths. The updated  transfer functions are then

\begin{equation}\label{eq:svar_equiv}
H_{hp}(z) =\frac {1}{1 + K/Q + K^2} \left[\frac {(1 - z^{-1})^2} {1 - \frac {(2(1 - K^2) z^{-1} -  (1 - K/Q + K^2) z^{-2}}{1 + K/Q + K^2}}\right],
\end{equation}
\smallskip

\noindent for the highpass output,

\begin{equation}
H_{bp}(z) = \frac {K}{1 + K/Q + K^2} \left[\frac {1 - z^{-2}} {1 - \frac {(2(1 - K^2) z^{-1} -  (1 - K/Q + K^2) z^{-2}}{1 + K/Q + K^2}}\right],
\end{equation}
\smallskip

\noindent for the bandpass output, and

\begin{equation}\label{eq:svar_fix_lp}
H_{lp}(z) = \frac {K^2}{1 + K/Q + K^2} \left[\frac {(1 + z^{-1})^2} {1 - \frac {(2(1 - K^2) z^{-1} -  (1 - K/Q + K^2) z^{-2}}{1 + K/Q + K^2}}\right],
\end{equation}
\smallskip

\noindent for the lowpass output. 

\subsection{Equivalence to Bilinear Transformation}

A cursory look at the numerator of these transfer functions indicates that we have zeros at $z = 1$ and $z = -1$, for the
bandpass case, and at $z=-1$ in the lowpass frequency response. The highpass transfer function keeps its two zeros
at $z=1$ as we should have expected. In fact these frequency responses are what we would expect if we were applying a bilinear transformation to the analog filter transfer functions. We can demonstrate this by setting $Q  = \sqrt{2}$, which should 
give the filter a Butterworth response. We can then compare to the classic definition of such a filter,
given by

\begin{equation}\label{eq:butt_lp}
H(s)H(-s) = \frac 1 {1 + (-s^2)^{N}},
\end{equation}
\smallskip

\noindent which describes a Butterworth response with a cutoff radian frequency $\Omega = 1$ \cite{Oppenheim:1999:DSP, Steiglitz}. 
For a second-order filter we set $N=2$. The poles of this filter are 

\begin{equation}\label{eq:bilinear_transform1}
1 + (-s^2)^{2}= 0,
\end{equation}
\smallskip

\noindent and there are four of these at the unit circle in the s-plane, whose
phases are $s_p = \{3\pi/4, -3\pi/4, -\pi/4, \pi/4\}$, of which only the first two are
stable. The locations of the poles in the z-plane are found using
the bilinear transformation,

\begin{equation}\label{eq:bilinear_transform_zeros}
z_p = \frac {1 + e^{\pm j 3\pi/4}}  {1 - e^{\pm j 3\pi/4}},
\end{equation}
\smallskip

\noindent which uses the conformal mapping $z = (1 + s)/(1 - s)$. We now use the
bilinear transformation in the other direction to obtain the 
transfer function of the digital filter,

\begin{equation}\label{eq:butt_lp_tf}
H(z)H(-z) = \frac 1 {1 + \left[-\left(\frac {z - 1} {z + 1}\right)^2\right]^N}.
\end{equation}
\smallskip

For $N=2$ we have,

\begin{equation}\label{eq:butt_lp_tf2}
H(z)H(-z) = \frac 1 {1 + \left(\frac {z - 1} {z + 1}\right)^4} =  \frac {(z + 1)^4} {(z + 1)^4 + (z - 1)^4},
\end{equation}
\smallskip

\noindent which shows that we also have four zeros, in addition to the four poles
shown above. These are all located at $z=-1$, which makes sense for a lowpass
filter. Since only the two first poles are stable and we want a second-order
filter, we will use two of these zeros in the final filter. 

From the digital transfer function, we can get the filter power spectrum,

\begin{equation}\label{eq:butt_lp_ps}
|H(\omega)|^2 = \frac 1 {1 + \left[-\left(\frac {e^{j\omega} - 1} {e^{j\omega} + 1}\right)^2\right]^N } = \frac 1 {1 + \tan^{2N}(\omega/2)}.
\end{equation}
\smallskip

\noindent The cutoff frequency of this filter satisfies $\tan^{2N}(\omega/2) = 1$, and so should be equivalent to
the digital state variable lowpass with $K=1$ and $Q = \sqrt{2}$. Replacing these parameters in
Eq.~\ref{eq:svar_fix_lp} demonstrates that this is indeed the case. In fact, we can now also see that
if we make the replacement $s = \frac {z - 1} {z + 1}$ in Eq.~\ref{eq:svar_an}, we will arrive
at a similar result to Eq.~\ref{eq:svar_hp_fix}. Another way to look at this is to say that 
it is also equivalent to applying the bilinear transform to the analog integrator 
transfer function $s^{-1}$. Therefore we have indirectly derived three
bilinear transformation digital filters, one for each of the three analog state variable responses.
As we noted earlier, we could of course use them to implement three separate filters using a digital 
biquadratic structure, but that is not our objective.

\subsection{Filter Equations}

We now need to apply the modifications from the transfer functions to implement an improved
state variable filter. This should follow from the recognition that each integrator should be
changed to include a zero at $z=-1$, 

\begin{equation}\label{eq:integtf2}
H(z) = \frac {1  + z^{-1}} {1 -  z^{-1}}.
\end{equation}
\smallskip

In fact, the equivalence to the bilinear transform, which we have noted earlier, is made fairly explicit in
this equation. Eq.~\ref{eq:integtf2} is equivalent to the bilinear transform (not normalised) \cite{STEIGLITZ1965}, 

\begin{equation}\label{eq:btrs}
s = \frac {z - 1} {z + 1},
\end{equation}
\smallskip

\noindent applied to the integrator transfer function $1/s$.

\begin{figure}[htp]
\begin{center}
\includegraphics[width=.4\textwidth]{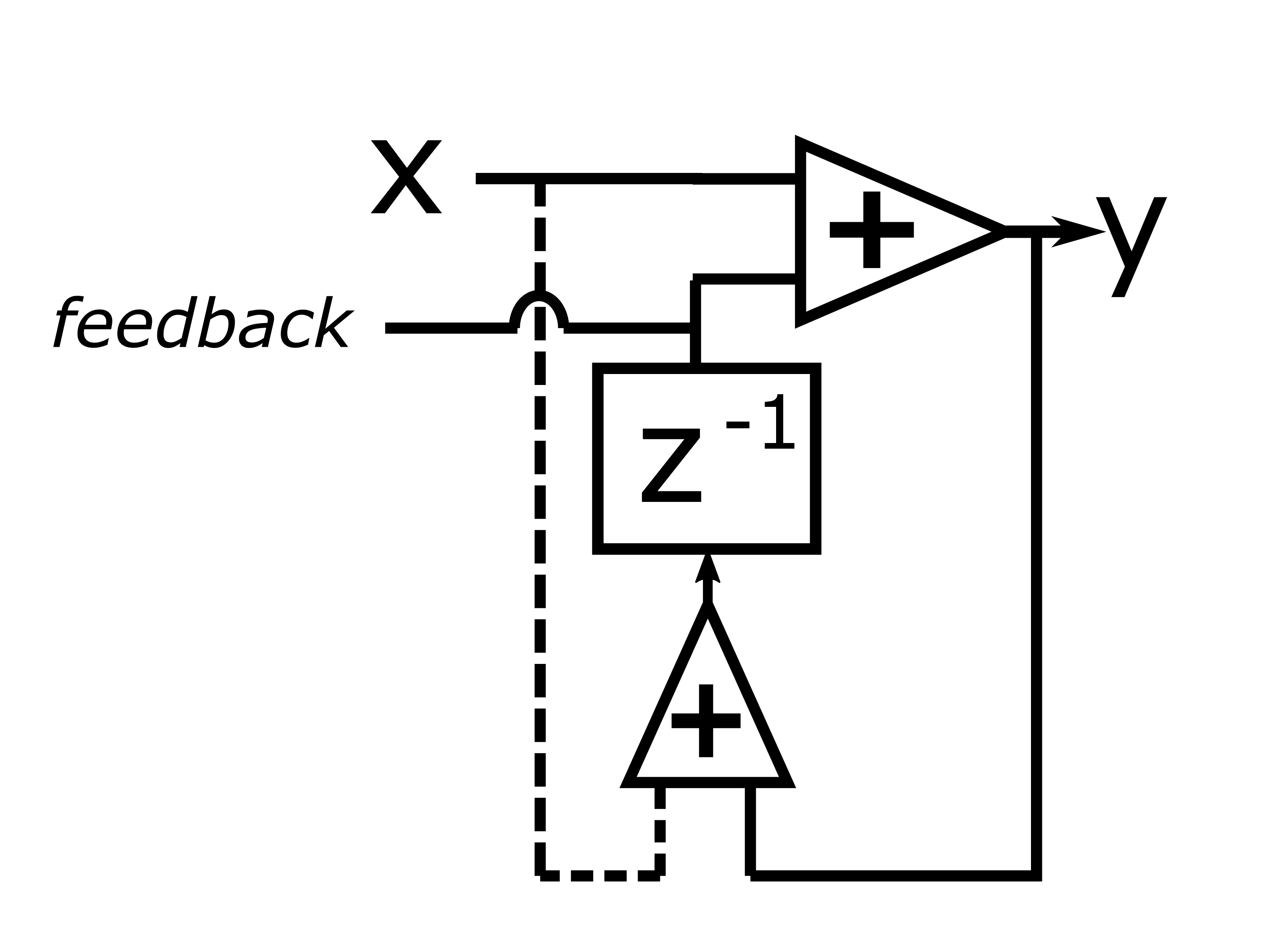}
\caption{The digital integrator with an added feedforward path.}
\label{fig:integrator2}       
\end{center}
\end{figure}

This change requires us now to change the integrator equation slightly.
Since it is important for us to continue to tap the filter state so to avoid an
extra 1-sample delay, we should implement the transfer function of Eq.~\ref{eq:integtf2} in such a way
which will allow us to preserve the feedback paths in the re-arranged Chamberlin 
design. As shown in Fig.~\ref{fig:integrator2}, we only need to add a feedforward path
to the allpole integrator to turn it into a 1-pole 1-zero configuration, as required by
the transfer function. For this we need to use a system of update equations,

\begin{equation}\label{eq:integtf3}
\begin{split}
&y(n) =  x(n) + s(n)  \\
&s(n) =  y(n) + x(n)
\end{split}
\end{equation}
\smallskip

\noindent where $s(n)$ now represents the filter delay (its state). The update order is
important here, as the filter state is only changing after its output sample 
has been produced.

From this, we can re-define the filter update equations as 

\begin{equation} \label{eq:svar_new}
\begin{split}
&y_{hp}(n) =  x(n) - (1/Q)s_{bp}(n) - s_{lp}(n) \\
&y_{bp}(n) = Ky_{hp}(n) + s_{bp}(n) \\
&s_{bp}(n) = y_{bp}(n)  +  Ky_{hp}(n) \\
&y_{lp}(n) = Ky_{bp}(n) + s_{lp}(n) \\
&s_{lp}(n) = y_{lp}(n)  +  Ky_{bp}(n).
\end{split}
\end{equation}
\smallskip

\subsection{Filter Stability}

We are almost finished, except for one aspect, which is filter stability. The filter continues to be
unstable as the original, particularly at higher frequencies. For this reason, the filter needs to be corrected 
so that it can be made stable within a wide range of values for $Q$ and $K$, or at least as stable as 
the derived transfer functions. Therefore we can derive the required adjustments by making the 
state variable filter frequency response equivalent to the derived biquadratic transfer function. 
At the moment, this is not the case. The main differences reside in the feedback paths to
the highpass output. While the state variable filter uses the previous first and second-order 
integrator states, the biquadratic transfer function expects that the actual integrator outputs,
$H_{bp}(z)$ and $H_{lp}(z)$, are used.  

We first need to recognise that as an input signal recirculates through the integrator state, it has 
a factor of two scaling with respect to the signal from the integrator output. We can use this 
as the basis for the derivation of a solution. First we will represent
the highpass signal output as $Y_{hp}$ and its input as $X$. The feedback signals in this case 
are formed by ${2KY_{hp} + S_{bp}}$ and ${2K(KY_{hp} + S_{bp}) + S_{lp}}$. The states $S_{bp}$ 
and $S_{lp}$ are associated with the first- and second-order integrators (which provide the 
bandpass and lowpass outputs). From this, we have for the state variable filter

\begin{equation}
\begin{split}
&Y_{hp} =  X - 2KY_{hp}/Q - S_{bp}/Q - 2K^2Y_{hp} - 2KS_{bp} - S_{lp} \\
&Y_{hp} =  \frac {X - KY_{hp}/Q - S_{bp}(1/Q + K) - K^2Y_{hp} - KS_{bp} - S_{lp}} {1 + 2K/Q + 2K^2}.
\end{split}
\end{equation}
\smallskip

With this result we have derived the corrections to make the filter behave in the same way
as the biquadratic form for which we have a transfer function. This is because we eliminated
the factors of two involved in the feedback signals, which was the difference between the filters. 
From this result, we conclude that to stabilise the filter, we need to do two things: \\

\begin{enumerate}
\item Scale the highpass output by $(1 + K/Q + K^2)^{-1}$; and 
\item Include the extra $-KS_{bp}$ term, which amounts to offsetting the first-order
feedback path gain $-1/Q$ by $-K$. 
\end{enumerate}
\bigskip

\begin{figure}[htp]
\begin{center}
\includegraphics[width=.4\textwidth]{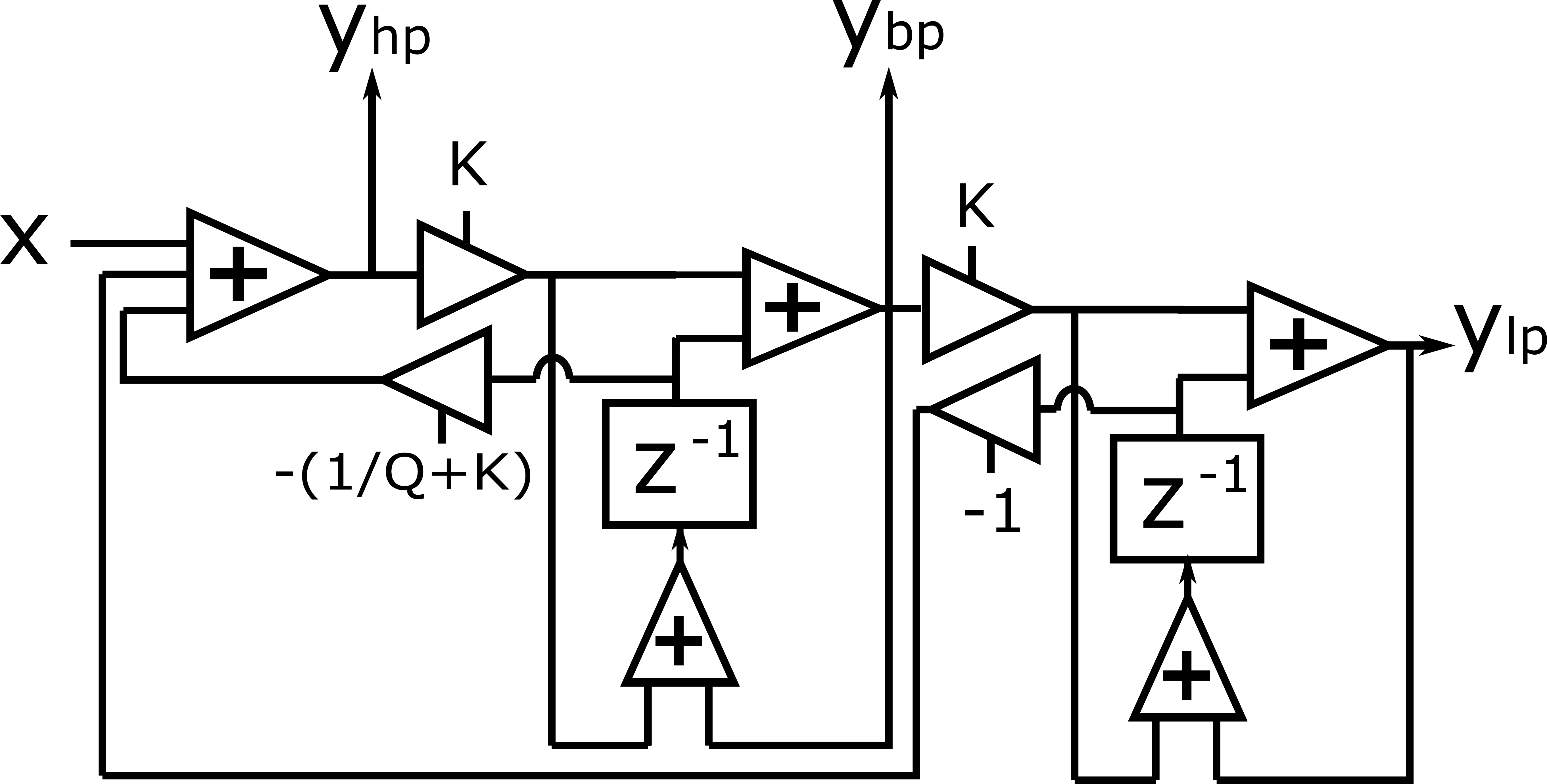}
\caption{Re-designed digital state variable filter block diagram.}
\label{fig:statevar3}       
\end{center}
\end{figure}

Without these corrections the feedback signals can render the filter numerically unstable. Note that 
these errors only become significant as $K$ gets larger, which is the case as the frequency 
increases. The $Q$ factor also plays a part in this, particularly if it is small. 
The final expression for highpass output is then

\begin{equation}\label{eq:svar_hp_correct}
y_{hp}(n) = \frac {x(n) -  (\frac 1 Q + K) s_{bp}(n) -  s_{lp}(n)} {1  + \frac K Q + K^2},
\end{equation}
\smallskip

\noindent which we can now replace in Eq.~\ref{eq:svar_new} to give the corrected filter equations.
The block diagram of this re-designed filter is shown in Fig.~\ref{fig:statevar3}.

\begin{figure}[htp]
\begin{center}
\includegraphics[width=\columnwidth]{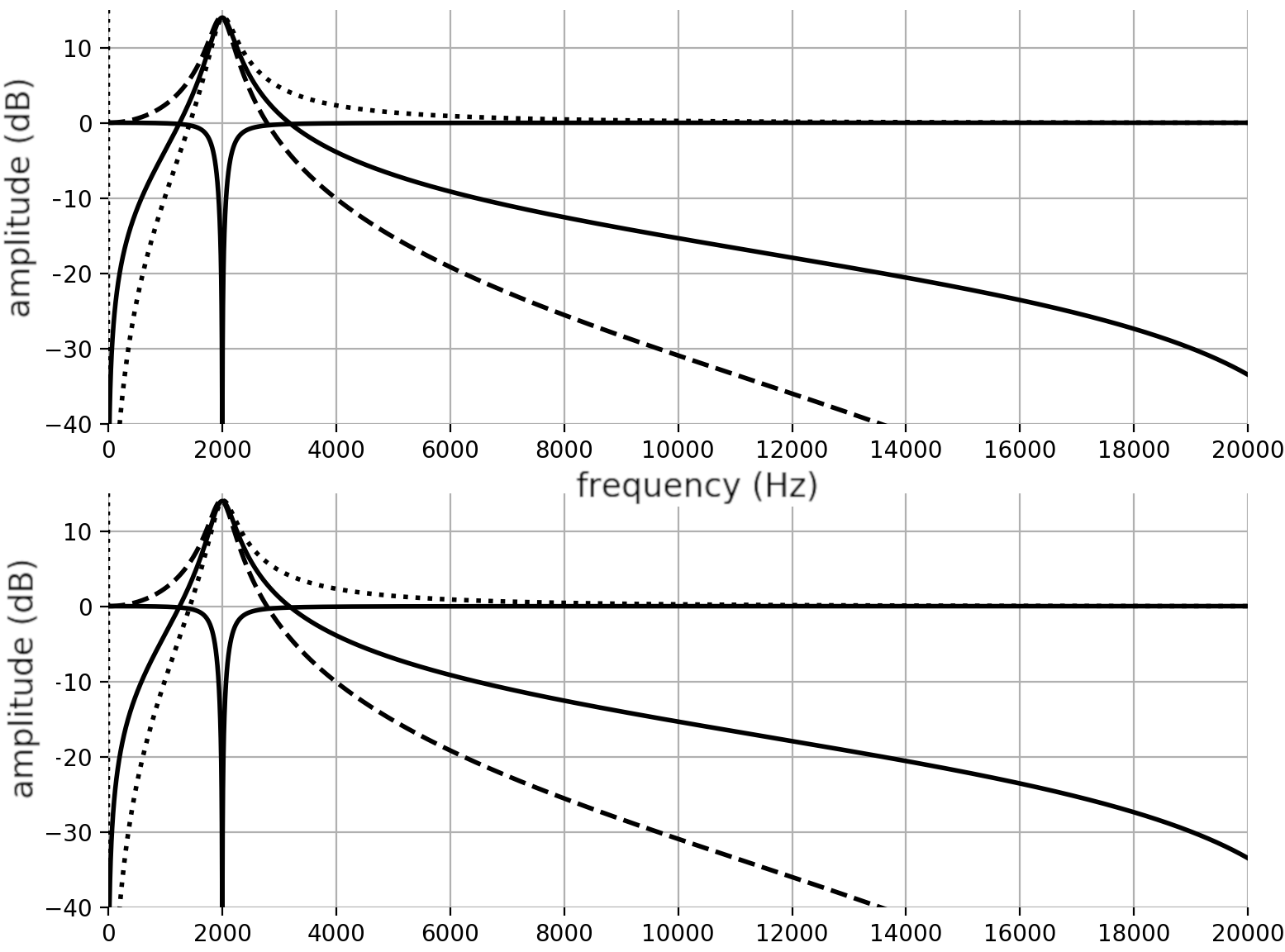}
\caption{Amplitude responses for the lowpass (dashes), bandpass(solid), band-reject (solid) and
highpass (dots) outputs of improved digital state variable filter (lower) and its equivalent biquadratic
filter transfer function (top).}
\label{fig:svardigi}       
\end{center}
\end{figure}

With these modifications, the transfer function of the three digital biquadratic filters and the state variable
frequency responses describe exactly the same amplitude spectrum. In Fig.~\ref{fig:svardigi}, 
we plot the biquadratic and state variable amplitude responses for the four basic filter outputs.
The first one was obtained by evaluating the transfer function directly, and the second from the
discrete Fourier Transform of the state variable filter impulse responses.

\subsection{Filter Frequency}

The digital state variable form developed here solves the difficulties with tuning we had
experienced with the Chamberlin model. However, we now need to find out
a different way to compute $K$ in such a way that a filter frequency parameter can be applied.  As
expected, one of the added bonuses of the method developed here is that 
now we have a filter whose transfer function has been warped correctly to fit within the digital 
baseband. This maps frequencies in such a way that the radian frequency $\Omega = \infty$ 
in the original analog filter is now $\Omega = \pi$ in this digital version. This can be easily
demonstrated by the fact that any zeros at $s = \pm \infty$ infinity are now placed at $z = -1$,
which is equivalent to the Nyquist frequency.

Thus all we need to do is to warp the filter frequency in the same way by applying a 
tangent map,

\begin{equation}
\Omega = \tan(\pi f/f_s),
\end{equation}\smallskip

\noindent and set $K = \Omega$. The filter is then good to go. Note that this is consistent with the fact that
the digital filter is now equivalent to one obtained through the application of a 
bilinear transformation to the analog state variable design.

\subsection{Band-Reject and Allpass Responses}

This filter allows us now to get the band-reject response as the sum $Y_{hp} + Y_{lp}$, 
or, alternatively, as $X - (1/Q + K)S_{bp}$, since these two expressions are equivalent.
A closer look at the filter equation will confirm that, as in the original analog filter, 
the highpass and lowpass responses are correctly offset by $\pi$ radians at their cutoff 
frequencies. 

The bandpass output, in the current form, does not have unity gain at the centre
frequency. However, it is a simple matter of scaling it by a $1/Q$ factor in order to
rectify this. As we can see, this is particularly useful if we want to make sure that
the output of the filter does not increase as we employ a sharper resonance. This
also provides the second form of the notch filter given earlier.

Finally, since we have both a phase-aligned band-reject and normalised bandpass outputs,
we can now obtain the allpass response that was missing from the Chamberlin filter. Due to
the extra delay between the three filter outputs, this was not possible to obtain directly. In
the current design, since we have restored the phase alignment in the original analog
filter, we can also get the allpass output as  $y_{hp} + y_{lp} + y_{bp}/Q$, which combine 
the opposing band-reject and normalised bandpass responses. Alternatively, the allpass
response can also be obtained by summing the input and twice the phase-inverted 
normalised bandpass output.

The complete filter with
the five outputs, highpass, lowpass, bandpass, band-reject, and allpass is shown in
Listing~\ref{code:svar3}.

\begin{lstlisting}[caption={Improved digital state variable filter.},label={code:svar3}]
opcode Svar3,aaaaa,akk
 setksmps 1
 as1,as2 init 0,0
 as,kK,kQ xin
 kdiv = 1+kK/kQ+kK*kK
 ahp = (as - (1/kQ+kK)*as1 - as2)/kdiv
 au = ahp*kK
 abp = au + as1
 as1 = au + abp
 au = abp*kK
 alp = au + as2
 as2 = au + alp
 xout ahp,abp,alp,
     ahp+alp,ahp+alp+(1/Q)*abp
endop
\end{lstlisting}

\section{Discussion}

\begin{figure}[htp]
\begin{center}
\includegraphics[width=\columnwidth]{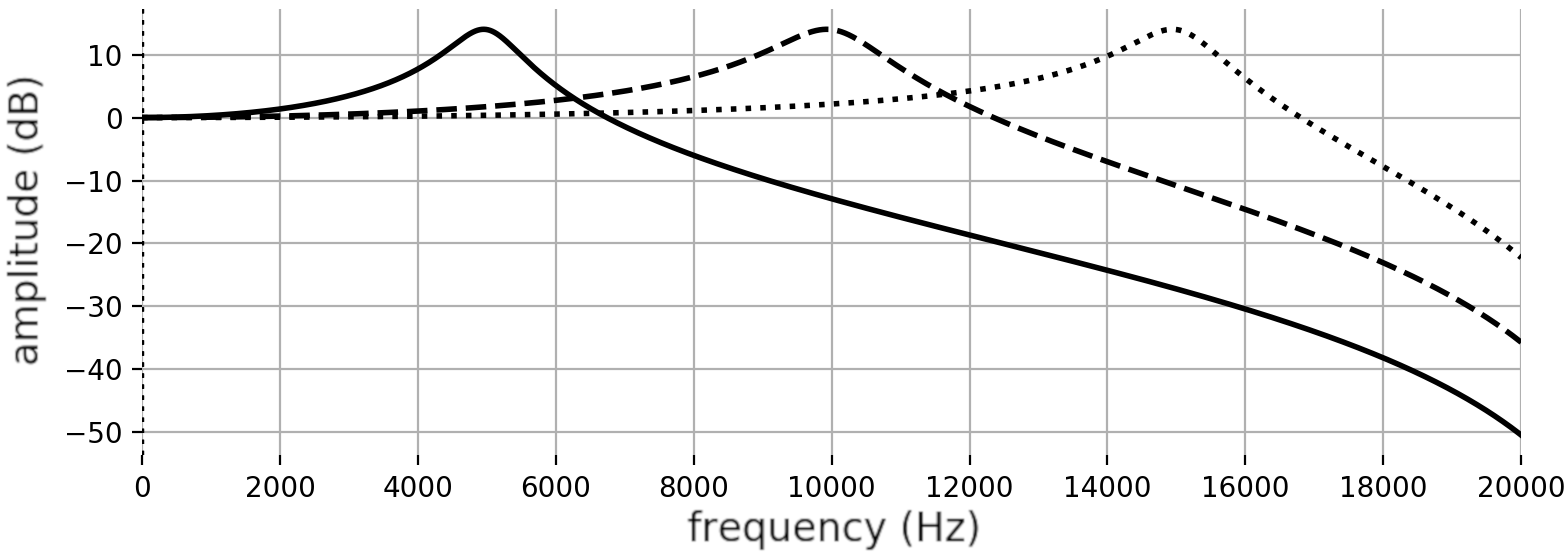}
\caption{Amplitude responses for the revised state variable lowpass filter, 
using Q=5 and $f=5$ KHz (solid), 10 KHz (dashes), and 15KHz (dots), with $K = \tan(\pi f/f_s)$ and $f_s =44.1$ KHz.}
\label{fig:svarfix}       
\end{center}
\end{figure}

As expected, the improved digital state variable filter has a much better high-frequency behaviour than
the Chamberlin design. Amplitude responses for the lowpass output are given in Fig.~\ref{fig:svarfix}.
These are now correct for a digital filter, with no high-frequency
issues at normal sampling rates. We can clearly see the beneficial effect of the zeros
we added to the integrators, as the high end of each curve is well anchored at the
Nyquist frequency (unlike in the previous case of Fig.~\ref{fig:svarerr}). Moreover,
the rearrangement of the filter equation also had separate effect on the position of
the poles, which can be surmised by looking at the differences between the denominators of the
transfer functions for the Chamberlin state variable filter and the improved version developed here.

Such changes can be explained by the two different methodologies of discretization that 
underline the digital filter models.  In the case of the Chamberlin design, the pole frequencies 
approximate the frequency of the analog filter poles, with an error that is inversely proportional to
the ratio of the sample rate and filter frequency. By increasing the sampling rate, we can minimise
the error, at the cost of extra computation. With extremely short sampling periods, the Chamberlin 
model will approximate the actual analog filter fairly well. This is also the case of the 
improved design, but with the usual sampling rates of $f_s=44.1$ to $48$ KHz, we have
a more reasonable warping of the frequency response. While both discretization methods inevitably lead 
to some distortion of the analog filter transfer function, the one in the improved design 
is of a more benign nature.

\subsection{Contrasting Approaches}

The method used to derive an improved filter was purely based on an analysis of the filter transfer functions,
which led to the development of a modified set of filter equations. However, it is possible to approach the 
problem from an alternative perspective, leading to exactly the same results. This starts by recognising that
the simultaneous nature of the three outputs in the analog filter is incompatible with a digital filter implementation.
As we already noted, there is something in the original filter that cannot be computed in a sequence of steps,
which is sometimes described as a \emph{delay-free loop}. The only way to deal with such a problem is
to introduce a 1-sample delay somewhere in the signal path, and we have shown that it matters where this
is placed. In some places, a design such as the one derived here is called a zero-delay filter, but that is a 
complete misnomer and such a term should be discouraged. While it is true that we placed the three outputs
in phase alignment, and as such any delays between them have been removed, it is impossible to completely 
remove feedback delays, and have instantaneous signals everywhere in the filter.  Instead, what we can do 
is move the 1-sample delays around, which gives the equations better resilience to errors.

\begin{figure}[htp]
\begin{center}
\includegraphics[width=.33\textwidth]{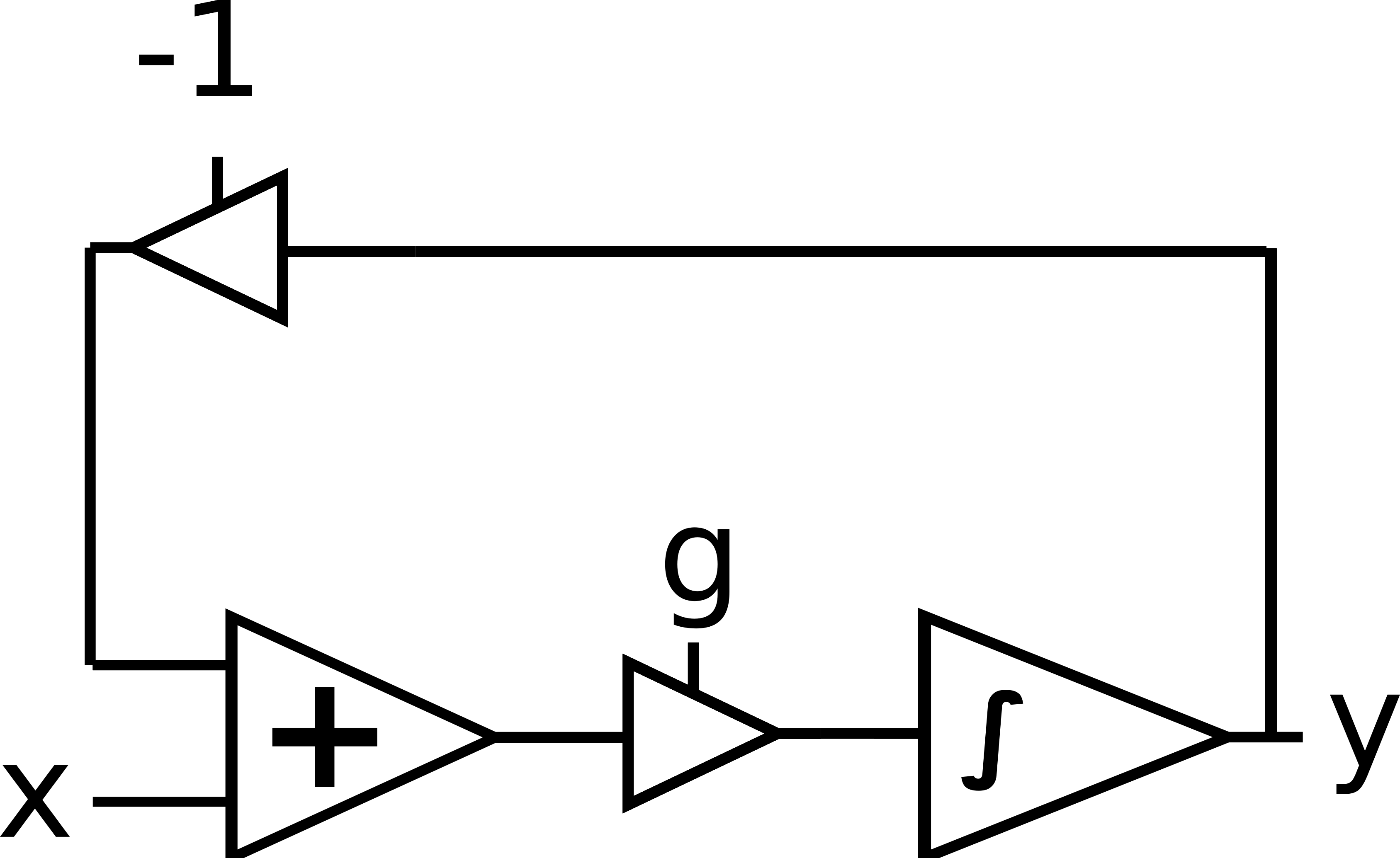}
\caption{First-order linear lowpass filter.}
\label{fig:linearlowpass}       
\end{center}
\end{figure}

We can demonstrate, however, that using a well-known algorithm for tackling delay-free loops \cite{Harma, Fontana2, Fontana1}, further developed by Fontana \cite{Fontana4}, and then D'Angelo \cite{Dangelo2} for non-linear cases, can lead to the exact same result we have obtained before. This approach involves no spectral domain considerations, it is purely focused on the rearrangement of the filter update equations.  In order to develop the idea, we first show how this can be applied to a leaky-integrator lowpass analog design, whose block diagram is shown in Fig.~\ref{fig:linearlowpass}. An example of such filter is given by the first-order sections in the Moog ladder filter \cite{Fontana1} ,

\begin{equation}\label{eq:mladder}
H(s) = \frac {\Omega} {\Omega + s},
\end{equation}\smallskip

\noindent as modelled for instance by Huovilainen \cite{Huovilainen} using the backward Euler method (but excluding the hyperbolic tangent non-linear mapping). 
The digital filter equation for this is given as

\begin{equation}\label{eq:firsto}
y(n) = g(x(n) - y(n-1)) + y(n-1).  
\end{equation}\smallskip

We first note that the $-gy(n-1)$ term on the right-hand side is, in this case, where a one-sample delay was inserted 
in order for the filter to work as a straight discretization of  the continuous-time differential equation. This eliminates a
delay-free loop in the analog filter flowchart, and the digital implementation follows from it. However, we can do better by 
being bold and defining what looks like a more correct model as

\begin{equation}
y(n) = g(x(n) - y(n)) + y(n-1).  
\end{equation}\smallskip

The remaining $y(n-1)$ term is the integration state, which we need to preserve. Note that, in this form, we have no
hope to compute its output, but we can proceed with the algorithm defined by H\"{a}rm\"{a} \cite{Harma} to get a usable set of
filter equations. Rewriting $s(n) = y(n-1)$ and re-arranging, we have

\begin{equation}
\begin{split}
&y(n) + gy(n) = gx(n) + s(n)  \\
&y(n)(1 + g) =  gx(n) + s(n)  \\
&y(n) = \frac {gx(n) + s(n)} {1 + g}.
\end{split}
\end{equation}\smallskip

The next step is to define a tap containing the signal before the integration stage, $u(n) = g(x(n) - y(n))$,
and do the replacement  

\begin{equation}
\begin{split}
&u(n) = g\left(x(n) - \frac {gx(n) + s(n)} {1 + g}\right)  \\
&u(n)=  g \frac {x(n) - s(n)} {1 + g}  .
\end{split}
\end{equation}\smallskip

What is left to do now is to order the operations carefully so that the filter output can be computed
correctly. Starting with $u(n)$, we need to obtain $y(n)$ first, then update the integration state,

\begin{equation} \label{eq:hama}
\begin{split}
&u(n) = g \frac {x(n) - s(n)} {1 + g}  \\
&y(n) =  u(n) +  s(n)\\ 
&s(n) =  y(n) +  u(n).
\end{split}
\end{equation}\smallskip

The relevance of this approach to our state variable problem can be demonstrated by showing
that the structure derived for the leaky integrator can be applied directly to modify the 
allpole integrator used in the Chamberlin filter. This can be shown as follows.
Starting with the filter

\begin{equation}
y(n) = Kx(n) + y(n-1)
\end{equation}\smallskip

We replace $y(n-1)$ by $s(n)$ and follow the steps outlined earlier.
From this, we will obtain the following update equations 

\begin{equation}
\begin{split}
&u(n) = Kx(n) \\
&y(n) =  u(n) +  s(n) \\ 
&s(n) =  y(n) +  u(n),
\end{split}
\end{equation}\smallskip

\noindent which gives us the integrator used in Eq.~\ref{eq:svar_new}. Conversely, we
can also apply our approach of integrator replacement to the first-order section of
Eq.~\ref{eq:firsto}. As we noted earlier, this is equivalent of placing a zero at the
Nyquist frequency ($z = -1$), which yields the following filter equation,

\begin{equation}\label{eq:firstor}
y(n) = g[x(n) + x(n-1)] - (g-1)y(n-1).  
\end{equation}\smallskip

\begin{figure}[htp]
\begin{center}
\includegraphics[width=\columnwidth]{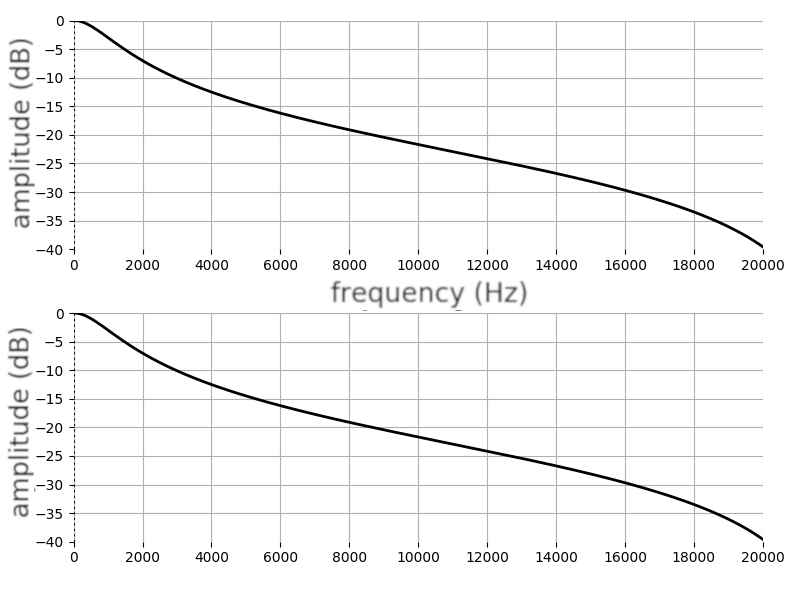}
\caption{Amplitude responses for two equivalent first-order lowpass filters constructed using H\"{a}rm\"{a}'s algorithm (Eq.~\ref{eq:hama}, top)
and our integrator replacement approach (Eq.~\ref{eq:firstor}, bottom) with $f_s =44100$  Hz and cutoff frequency $f_c=1000$ Hz. These figures
were generated from the impulse responses of the respective filters.}
\label{fig:firsto}       
\end{center}
\end{figure}

As we have done earlier for the state variable filter, we need to correct it so that the effects
of the changed integrator are accounted for. For this we simply need to scale  the equation by $(g+1)^{-1}$,
and we have a filter that is equivalent to the one in Eq.~\ref{eq:hama}. The transfer function is

\begin{equation}\label{eq:firstortf}
H(z) = \frac g {g + 1}  \left( \frac { 1 + z^{-1} } { 1 + \frac{g-1} {g+1} z^{-1}} \right),
\end{equation}\smallskip

\noindent with typical amplitude responses as shown in the plots of Fig~\ref{fig:firsto}. 
This completes the proof that our approach matches H\"{a}rm\"{a}'s method of 
delay-free loop elimination. We should also note that it is possible to arrive at Eq.~\ref{eq:firstortf} via
a third route, which is to apply the bilinear transform (Eq.~\ref{eq:btrs}) to the analog filter transfer function (Eq.~\ref{eq:mladder}).

While this technique is a well-established way to obtain better filters, particularly with regards
to tuning and high-frequency behaviour, it obscures the fact that the solution is leveraged 
by the anchoring of the transfer function at the Nyquist frequency.
Our original motivation for such a modification to the integrators was, on the other hand, 
a purely spectral one, which followed directly from the recognition that the theoretical zeros placed at the 
origin in the lowpass case were not at an optimal location. This led to the incorporation of the 
one-sample feedforward  delay into the integrator, which was done in line with the aims of 
H\"{a}rm\"{a}'s algorithm, that is, to avoid the introduction of an extra delay in the filter 
update equations. 

\section{Conclusions}

In this article, we have looked at the state variable filter and its typical digital implementation given by Chamberlin \cite{Chamberlin},  and proposed some modifications leading to an improved frequency response at the critical sampling rates for full-band audio (e.g. 44.1 or 48 KHz). With it, it is possible to preserve the original filter block diagram, which allows  us to compute four simultaneous frequency responses with a small number of operations. We have also noted that the spectral method developed here effectively targeted the transformation of two zeros at infinity of the s-plane into zeros at the Nyquist frequency point of the z-plane, one in each integrator. This played an important part in correcting the amplitude responses, particularly for the bandpass and lowpass cases at high filter frequencies. We then demonstrated that this yields a state-variable filter that is equivalent to one obtained by applying the bilinear transformation to the analog transfer function. This result proves that it is possible to employ the bilinear transformation to obtain a digital version of the state variable filter using a similar topology to the analog case.

This method was compared with the well-known approach of re-arranging filter equations to tackle the issue of delay-free loops, and we concluded that the two  alternative approaches can lead to the same results in the present case. Finally, it is important to note that the original design by Chamberlin is correct, and will work well if the sampling frequency is sufficiently high since with a relatively small unit delay, the original analog filter will be well approximated. In contrast, the filters derived here have a warped frequency response that is not equivalent to the analog case, but are less correct from that perspective. On the other hand, they can produce better results at lower sampling rates.

The filter we have arrived at, with its warped frequency response, is therefore a practical compromise (as all bilinear-transform filters are), which works very well in applications where significant oversampling is not desirable. Along with the original Chamberlin design, its main virtue is that it preserves a topology that resembles very closely its analog counterpart. This is particularly useful in situations where we aim to go beyond the linear operation to embrace various types of nonlinear distortion, as we can model these by following the location of the sources of nonlinearities in the analog filter \cite{CMJ}.

\bibliography{paper}
\bibliographystyle{aes2e.bst}

 \biography{Victor Lazzarini}{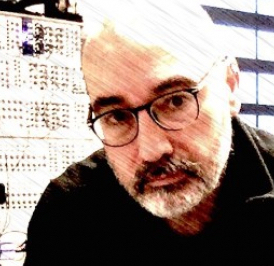}{Victor Lazzarini is Professor of Music at Maynooth University. He is a graduate of the Universidade Estadual de Campinas (UNICAMP) in Brazil, and completed his doctorate at the University of Nottingham, UK (1996). His research is focused on Computer Music Languages and Musical Signal Processing. Recent publications include Ubiquitous Music Ecologies (with D. Keller, N. Otero, and L. Turchet, 2020) and Spectral Music Design: A Computational Approach (Oxford Univ. Press, 2021).}
 \biography{Joseph Lazzarini}{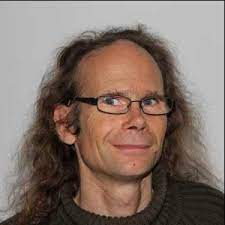}{Joe Timoney is Head of the Computer Science at Maynooth University. He studied Electronic Engineering, completing his PhD in 1998. He joined the Dept. of Computer Science at NUI Maynooth in the following year. He teaches on undergraduate programs in Computer Science and in Music Technology. His research interests are based in the area of audio signal processing, with a focus on musical sound synthesis and the digital modelling of analogue subtractive synthesis.}

\end{document}